# Cantilever-enhanced photoacoustic measurement of light-absorbing aerosols

Juho Karhu[a,*], Joel Kuula[b], Aki Virkkula[b], Hilkka Timonen[b], Markku Vainio[c,d], and Tuomas Hieta[e]

[a]Metrology Research Institute, Aalto University, Espoo, Finland; [b]Atmospheric Composition Research, Finnish Meteorological Institute, Helsinki, Finland; [c]Department of Chemistry, University of Helsinki, Helsinki, Finland; [d]Photonics Laboratory, Physics Unit, Tampere University, Tampere, Finland; [e]Gasera Ltd., Turku, Finland

*juho.j.karhu@aalto.fi

**ABSTRACT**

Photoacoustic detection is a sensitive method for measurement of light-absorbing particles directly in the aerosol phase. In this article, we demonstrate a new sensitive technique for photoacoustic aerosol absorption measurements using a cantilever microphone for the detection of the photoacoustic signal. Compared to conventional diaphragm microphones, a cantilever offers increased sensitivity by up to two orders of magnitude. Here we reached a noise level of 0.013 Mm$^{-1}$ (one standard deviation) with a sampling time of 20 s, using a simple single-pass design without a need for a resonant acoustic cell. We demonstrate the method in measurements of size-selected nigrosin particles and ambient black carbon. Due to the exceptional sensitivity, the technique shows great potential for applications where low detection limits are required, for example size-selected absorption measurements and black carbon detection in ultra clean environments.

## Introduction

Light absorbing particles, such as black carbon (BC), have an important role in the climate system of Earth. These particles absorb solar radiation and thus warm the surrounding climate (Bond et al. 2013). Moreover, they influence the optical properties of clouds and alter the albedo and thus the melting of snow and ice (Qian et al. 2015). In comparison to carbon dioxide, the lifespan of light-absorbing particles is relatively short, and therefore sustained reduction in their emissions represents a potential mitigation strategy for the alleviation of climate change in the short term (Xu and Ramanathan 2017).

To date, there are several different techniques, which can be used to measure the particle light absorption. A common method and an instrument is an aethalometer, which operates by depositing the sampled aerosol on a filter tape. The amount of collected particles is determined by measuring light transmittance through the filter. However, despite its wide use, the filter-mediated measurement technique introduces uncertainties, such as multiple-scattering and loading effect, which must be accounted for (Virkkula et al. 2007). In these sources of error, the fibrous material of the filter and the light absorbing particles gradually accumulating on the filter increase and decrease, respectively, the optical pathway and thus alter the measurement of light absorption. Moreover, the method requires intermittent replacement of the filter tape and a relatively long integration time to reach high sensitivity. An alternative approach to the filter-mediated aethalometer measurement technique is photoacoustic (PA) detection (Miklós, Hess, and Bozóki 2001). Photoacoustic detection is based on molecules absorbing energy of light illuminated towards them. When the material absorbs energy, it heats up and expands, which increases the pressure in an enclosed space. When the radiation is modulated with a certain frequency, the pressure variation within the photoacoustic sample chamber creates an acoustic wave of the same frequency. The light energy is converted into pressure variations i.e. sound energy, which is then converted into electric signal using a microphone. Although the technique is more commonly used in gas-phase measurements, it is applicable to particle-phased measurements as well (Petzold and Niessner 1995; Arnott et al. 1999). The greatest benefit of the PA technique is that the optical absorption measurement can be performed without the physical disturbance of the sample.

Recent developments in PA research have led to novel detection schemes, where the conventional microphone is replaced, for example, with an extremely sensitive miniature silicon cantilever. Cantilever-enhanced photo-acoustic spectroscopy (CEPAS) has been applied to several highly sensitive trace gas detection schemes from the visible to the mid-infrared wavelength regions; for instance, a noise equivalent detection limit ($1\sigma$) of 50 ppt in 1 s integration time was achieved in $NO_2$ detection using a visible light source (Peltola, Hieta, and Vainio 2015). The corresponding normalized noise equivalent absorption

(NNEA) coefficient was $2.6\times10^{-10}$ W cm$^{-1}$ Hz$^{-½}$. In this study, we describe a new measurement system where the CEPAS technique is extended to optical absorption measurement of particulate matter. The performance of the novel CEPAS aerosol sensor is evaluated in both laboratory and field experiments and compared to the existing instrumentation used in optical absorption measurements. A noise level of 0.04 Mm$^{-1}$ is reached with integration time of 1 s. This is, to the best of our knowledge, the best reported detection limit for a photoacoustic particle measurement and an order of magnitude improvement to a typical performance (for a concise summary of the sensitivities in recent PA measurements, see the supplementary material in Fischer and Smith 2018). The suitability of the CEPAS for various applications is discussed.

**Experimental materials and methods**

*Cantilever-enhanced photoacoustic spectroscopy*

In CEPAS, the microphone recording the photoacoustic signal is replaced with a micromachined silicon cantilever (Kauppinen et al. 2004). The dimensions of the cantilever are typically a few mm in width and length and less than 10 µm in thickness. The cantilever responds to the photoacoustic wave generated inside the sample cell by bending, and not by stretching like a typical diaphragm microphone. This allows much larger linear motion and leads to about two orders of magnitude enhancement in photoacoustic detection sensitivity. The movement of the cantilever is recorded with optical interferometry. Cantilever-enhanced PA detection has been used for gas phase absorption measurements ranging from visible wavelengths (Peltola, Hieta, and Vainio 2015) up to 15 µm in the infrared region (Karhu et al. 2020), as well as for measurements of optical radiation from ultraviolet to far infrared using carbon-soot absorbers (Rossi et al. 2021). Detection limits down to parts-per-trillion levels and below have been reached for various gaseous compounds (Peltola, Hieta, and Vainio 2015; Tomberg et al. 2018). For these measurement setups, the corresponding noise equivalent absorption coefficients, with 1 s measurement time, are typically around $5\times10^{-10}$ cm$^{-1}$ = 0.05 Mm$^{-1}$.

The CEPAS cell consists of the optical interaction volume and a balance volume, with the cantilever microphone located between the two (Kuusela and Kauppinen 2007). The interaction volume is a 4 mm in diameter and 90 mm in length hollow cylinder, through which a laser beam is shone. The interaction volume is enclosed on both ends with anti-reflection coated fused silica windows, at a slight angle to counteract etalon effects. At each end of the cylinder, voltage-controlled valves control gas flow into the CEPAS cell. The cantilever is located at the side of the cylinder, near the midpoint of the cylinder length. The total internal volume is about 8 ml.

*Laboratory evaluation of CEPAS with absorbing particles*

The novel CEPAS absorption measurement system was first tested and calibrated using nigrosin particles. Nigrosin is a water-soluble black dye, which absorbs strongly throughout the visible wavelength range. Aerosolized nigrosin has been used to validate various photoacoustic aerosol measurement systems (Lack et al. 2006; Davies et al. 2018; Fischer and Smith 2018; Wiegand, Mathews, and Smith 2014). Our measurement setup consisted of size-selected particle generation, CEPAS measurement system (model PA201, Gasera Oy, Finland) and a reference measurement with condensation particle counter (CPC; model 3776, TSI Inc., USA) (see figure 1). The particles were generated from a 5 mM nigrosin water solution with an atomizer (model ATM 226 Aerosol generator, Topas GmbH., Germany) and size-selected with a differential mobility analyzer (DMA; model 3081, TSI Inc., USA). The DMA sheet flow was set to 4 l/min. The particle flow between the atomizer and DMA was divided into two branches, which were reconnected before the DMA. One of the branches was sent through a particle filter and the flow through the unfiltered pipe was controlled with a variable flow valve. Partially closing the valve allowed for decreasing the total particle number in the reconnected flow arriving at the DMA. After the size-selection, the flow was divided symmetrically into two, with one branch going to the CEPAS system and the other to the CPC. The flow through the CPC was 0.3 l/min, and the flow rate through the CEPAS system was matched to the same value with a pump connected to the CEPAS system exhaust.

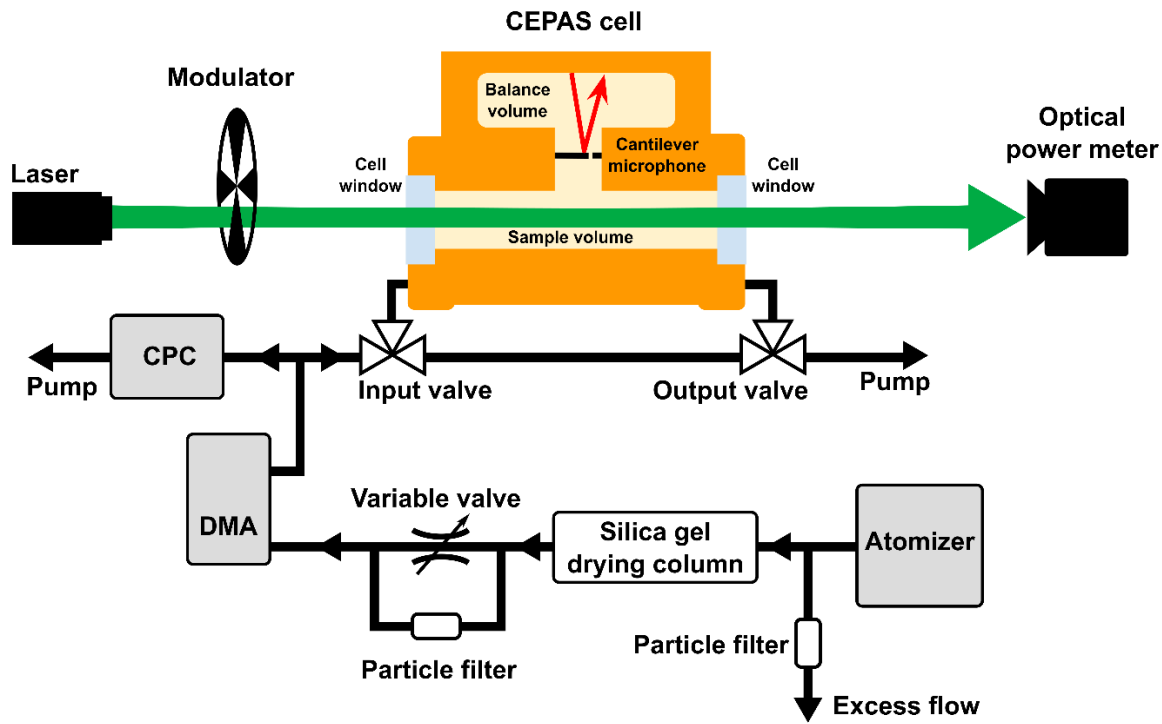

Figure 1: Schematic picture of the measurement setup, depicting the CEPAS measurement system and the sample line used in the laboratory evaluation measurements. The red arrow refers to the interferometric readout of the cantilever microphone. DMA: differential mobility analyzer; CPC: condensation particle counter.

The CEPAS measurement system was set up so that there was a constant flow of 0.3 l/min either through the CEPAS cell or pass it. During one measurement cycle, the flow was first directed through the cell for 10 s. Afterwards, the flow was switched to bypass the cell and the cell valves were closed for the measurement period. The pressure inside the cell was recorded when the valves were closed, and it was about 990 mbar. One sample was measured for 10 seconds and then the cycle was repeated. The light source in the CEPAS measurement was a high-power continuous-wave solid-state laser, emitting at 532 nm (model Verdi V-10, Coherent). The CEPAS cell was enclosed with fused silica windows, which were antireflection coated for the laser wavelength (Laser V-Coat, R < 0.25 % at 532 nm, Edmund Optics). The laser beam was amplitude modulated with a rotating chopper blade at a frequency of 85 Hz. The optical power after the CEPAS cell was monitored with a power meter (1918-C, Newport). Note that the laser beam was amplitude modulated with a 50% duty cycle, so the reported optical powers in the sample are approximately half of the power emitted from the laser. The signals from CPC, CEPAS and the power meter were collected with a computer running a LabVIEW measurement program. The CEPAS signal was phase-coherently demodulated digitally. The photoacoustic signal was normalized by calculating the ratio between the demodulated CEPAS signal and the measured optical power.

*Field evaluation of CEPAS*

The instrument was also tested with a three-day measurement of ambient air. Instead of the size-selected particle generation, the sample input was connected outside the laboratory with copper tubing. Prior the instrument, the air flow was dried with a silica column. The sample flow was divided between the CEPAS cell and the CPC to verify that the particles pass through the sampling line. The diffusion losses from the sampling pipeline and the impact losses from the 90° bend, which divided the flow into the CEPAS and CPC instruments, were estimated and the penetration efficiency is shown in figure S1 in the supplementary material. The results of the CEPAS measurement were referenced against an aethalometer (AE33, Magee Scientific Co., USA) located at SMEARIII air quality station (Järvi et al. 2009) a few hundred meters away (see figure S2 in supplementary material). The area in which the measurements were conducted is surrounded by a few public space buildings belonging to University of Helsinki and Finnish Meteorological Institute and low vegetation. The nearest major road, which is also the main local source of BC and $NO_2$, is located approximately 100 m to the east. The mass absorption coefficient for black carbon in atmospheric measurement can vary due to changes in the chemical composition and coating of the carbon particles, but it can be expected to be between about 7–12 $m^2/g$ (Bond, Habib, and Bergstrom 2006). A mass absorption coefficient of 10 $m^2/g$ was used in the analysis here, to convert the absorption results into black carbon mass concentration. Every few hours during working hours, the input flow was directed through a particle filter for 5 minutes to check the background level. A linear interpolation between the blank measurements was used to correct the background level in the concentration data.

## Results and discussion

*Laboratory evaluation of CEPAS*

The CEPAS system was first used to measure a gas sample with known concentration of $NO_2$, to calibrate the relation between the photoacoustic signal and optical absorption within the photoacoustic cell. The $NO_2$ absorption cross section at this wavelength is $1.5 \times 10^{-19}$ $cm^2$ $molecule^{-1}$ and the wavelength is mostly free of absorption from any interfering species in air (Peltola, Hieta, and Vainio 2015). The same measurement cycle was used as with the particle measurement. The optical power reaching the CEPAS cell was 1 W. The sample was from a gas cylinder with 10 ppm of $NO_2$ in nitrogen, which was diluted with nitrogen using mass flow controllers. The cell was flushed with the sample, until a stable photoacoustic signal level was reached. The CEPAS signal was measured from four $NO_2$ concentration levels, ranging between 208-521 parts-per-billion, with known absorptions. A conversion factor between the power normalized CEPAS signal and absorption was calculated with a linear least-squares fit to the $NO_2$ calibration set.

The noise performance of the CEPAS measurement was verified with a measurement of nitrogen filled cell. The signal thus corresponds to the background generated from residual absorption at the cell windows or by absorption of scattered light reaching the cell walls. The sampling rate was 1 s, and no gas exchange was applied in the noise test. The measurement was performed with optical powers of 0.5 W and 1 W. The CEPAS signal was converted to absorption according to the $NO_2$ calibration measurement and the Allan deviations of the absorption background signals from the two measurements are shown in figure 2. The higher optical power led to better noise performance at short time scale, but the drifts at longer averaging times start to become apparent. The noise level with optical power of 1 W and averaging time of 1 s was 0.04 $Mm^{-1}$, corresponding to an NNEA of $4 \times 10^{-10}$ W $cm^{-1}$ $Hz^{-½}$, which agrees with previous CEPAS measurements (Peltola, Hieta, and Vainio 2015). With 10 s averaging, which is the time used to measure each sample in the measurement cycle, the noise level was 0.013 $Mm^{-1}$.

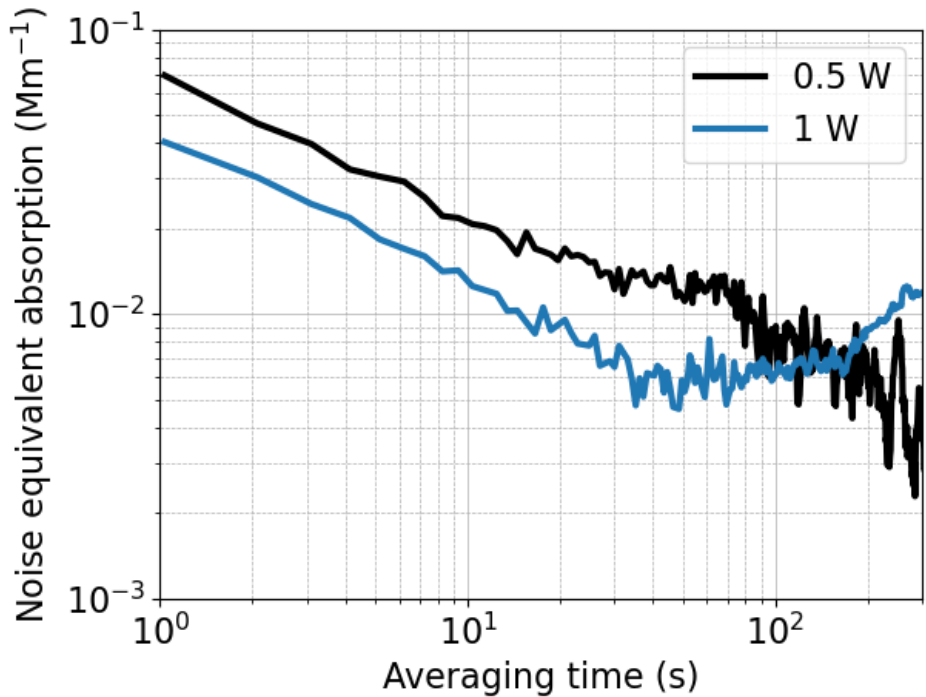

Figure 2: Allan deviation of CEPAS noise measurement. Measurement was performed with a nitrogen filled cell at two optical powers. With 0.5 W optical power, the signal was recorded for 30 min and with 1 W power for 15 min.

Response of the CEPAS system to absorbing aerosols was validated with nigrosin particles of diameter 100 nm, 200 nm and 300 nm. For each particle size, four concentration levels were measured. The particle concentration was varied with the valve controlling the flow bypassing the particle filter before the DMA. Optical power reaching the cell was 2.5 W. For each particle concentration level, the absorption was averaged for 10 minutes. Between each concentration step, 10 minutes of blank measurement was recorded. The average between the blank measurements on both sides of a sample was used to correct for a background level due to absorption from cell windows or walls. To generate the blank measurements, the flow from the atomizer was fully directed through the particle filter. The results of the concentration measurements are presented in figure 3. The absorption is the average over the 10-minute measurement and the error bars show the standard deviation over the averaging period. The number concentration is an average of the concentration recorded with the CPC over the same measurement time. The standard deviation of the concentration step with lowest measured absorption, corresponding to a number concentration of 44 cm$^{-3}$ with a diameter of 100 nm, was 0.053 Mm$^{-1}$. The figure 3 also shows a line fitted to each concentration set with a linear-least squares fit.

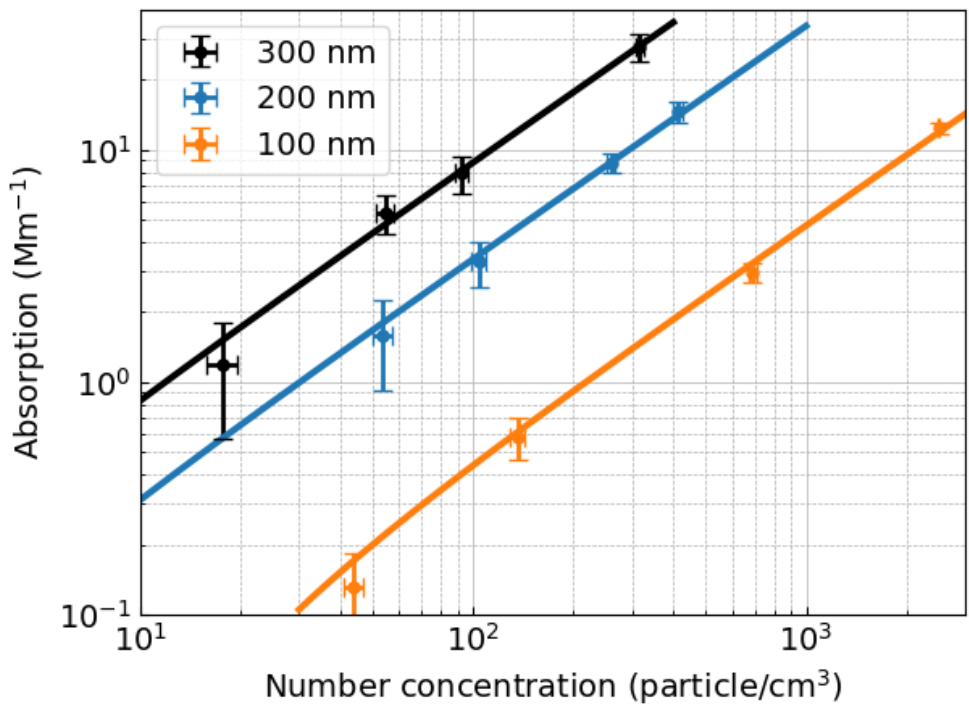

Figure 3: Nigrosin absorption signal recorded with CEPAS at different particle concentrations. The number concentration was measured with a CPC. Each point is the average over a 10-minute measurement and the error bars are one standard deviations of the measurement set.

*Ambient measurements*

The results of the outdoor air test measurement performed with CEPAS and the corresponding aethalometer data are shown in figure 4. The outdoor measurement was performed with optical power of 1 W. In the figure data, each data point is an average calculated over 1 hour period (n=167 for the CEPAS data and n=60 for the aethalometer data over 1 hour). The results show good agreement (r=0.91) between the two measurements over the three-day period, although some local variation is clearly present. It should be noted that in this test measurement, switching to the background measurement was done manually and only performed regularly during the daytime, which increases the uncertainty of the concentration measured with CEPAS during the nights. Overall, the measured BC concentrations were low during the test period; the long-term average in the urban background areas of Helsinki is approximately 520 ng m$^{-3}$ (Luoma et al. 2021) whereas in this study the average concentration was 97 ng m$^{-3}$ according to the aethalometer data. As expected, this seemed to have no hindering effect on the CEPAS performance.

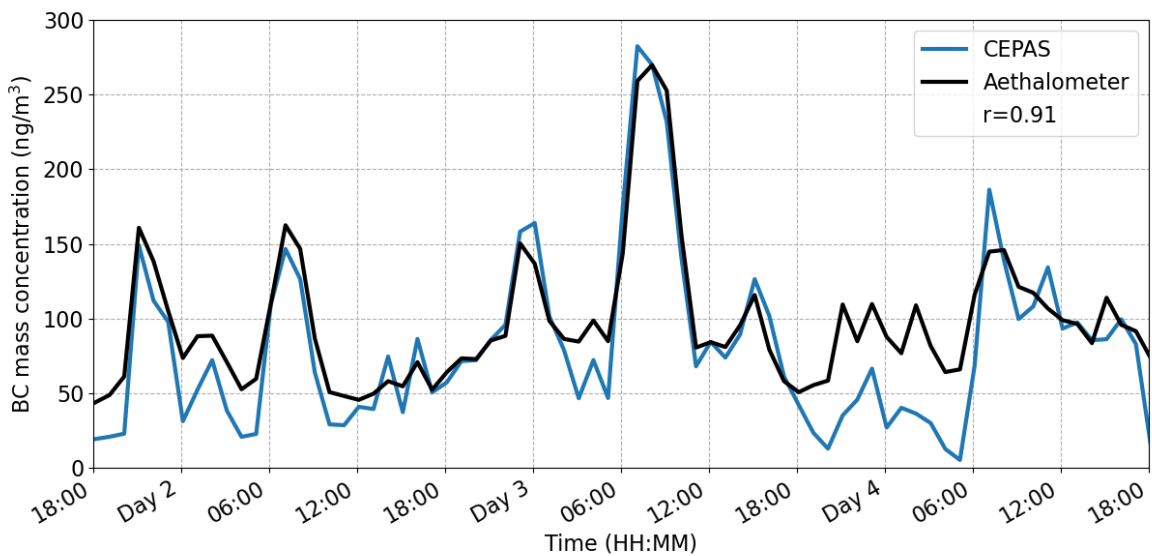

Figure 4: Outside test measurement. Each data point is an average over 1 hour in both datasets. The legend shows the Pearson correlation coefficient between the two datasets.

As noted in the laboratory evaluation, $NO_2$ has an appreciable absorption cross section at the wavelength of the laser used in the experiment. In fact, ambient concentrations should produce a photoacoustic signal of similar magnitude to that produced by the measured concentrations of black carbon aerosols. However, interference from $NO_2$ was not observed in our outdoor air measurement. To explain this observation, throughput of $NO_2$ through components of the sampling line was tested with a photoacoustic measurement. Constant flow with 1 ppm of $NO_2$, which was prepared the same way as described for the gas phase calibration, was sent through the filter used to correct the background and separately through the silica gel dryer column. Measurement with the sample flow going through the silica gel produced approximately 87% lower photoacoustic signal compared to a signal from uninhibited flow. This would have reduced the $NO_2$ concentrations reaching the CEPAS cell to insignificant levels in the outdoor measurement. Furthermore, the particle filter had no effect on the $NO_2$ photoacoustic signal. Therefore, the average signal from the residual $NO_2$ is included in the background subtraction.

**Conclusions**

Photoacoustic spectroscopy is a direct method for measuring optical absorption and thus its validity in particle measurements is not hindered by the typical uncertainty factors present in aethalometer-type instruments. Photoacoustic spectroscopy can be used to differentiate between the absorption enhancement of particles due to lensing effect of coated black carbon particles and the absorption contribution of brown carbon by utilising thermally denuded sample lines. For these reasons, photoacoustic spectroscopy has become a commonly used measurement technique in the atmospheric light absorption studies.

Conventional photoacoustic spectrometers rely on diaphragm microphones to detect and measure the photoacoustic signal generated in the sample cell. The cantilever-enhanced detector presented here responds to the photoacoustic signal by bending rather than stretching and thus surpasses the conventional microphone-based spectrometers in terms of sensitivity and linearity; the noise level of CEPAS reaches 0.04 $Mm^{-1}$ (1σ) with 1 s averaging time. With the 10 second averaging time used in the measurement cycle, the noise level is 0.013 $Mm^{-1}$ and when this is normalized to the total time of the sampling cycle (20 s), we get 0.058 $Mm^{-1}$ $Hz^{-½}$. This is over an order of magnitude better than most previously demonstrated photoacoustic aerosol absorption measurement techniques (Nakayama et al. 2015; Cao et al. 2021; Lack et al. 2006; Wang et al. 2017; Yu et al. 2019; Fischer and Smith 2018). The short-term noise could be further improved with higher laser power, but in our test measurements, the drifts due to the thermal management of the high-power laser started to limit the practical noise level at longer time scales. To the best of our knowledge, the lowest noise level previously reported was 0.15 $Mm^{-1}$ (2σ in 1 s), which was reached with a resonant cell and a multi-pass setup with up to 182 passes (Lack et al. 2012). It should be noted that the high sensitivity in our measurement is reaches with a simple single-pass configuration and the sensitivity could be further improved with a multi-pass measurement. Increasing the performance with a multi-pass setup instead of higher laser power would also require less demanding thermal management from the laser, which could lead to better long-term stability as well.

The CEPAS was further validated in a field test, in which a comparable result to that of the reference aethalometer was achieved. When considering field experiments, addition of more detection wavelengths is an important task and feature to be implemented as it enables a much more in-depth analysis of the particle light absorption properties. The technical implementation of a multi-wavelength detection is straightforward; the CEPAS does not operate at an acoustic resonant frequency, and therefore multiple wavelengths can easily be multiplexed to different modulation frequencies for simultaneous detection. In configuration relying on acoustic resonances, simultaneous measurement of multiple wavelengths requires more complex designs, such as multiple parallel PA cells (Ajtai et al. 2010) or acoustic cells which can accommodate multiple resonances (Cao et al. 2021). For a limited number of wavelengths, the modulation frequencies can be multiplexed over the acoustic resonance peak with small frequency differences (Yu et al. 2019; Fischer and Smith 2018), but this results in reduced sensitivity since the configuration does not take full advantage of the resonance (Cao et al. 2021) and the wavelengths further away from the true resonance frequency become more sensitive to environmental effects (Fischer and Smith 2018).

Taking into consideration the characteristic features of CEPAS, such as exceptional sensitivity, easy wavelength multiplexing and large dynamic range, several potential applications can be presented:

i) Light absorption measurements in ultra clean environments, such as in the Arctic and Antarctic regions, where black carbon plays an important role with respect to climate change, can be conducted at a higher accuracy than previously possible. This is a direct consequence of the high sensitivity of the CEPAS.

ii) When coupled with a Differential Mobility Analyzer, the fast and highly sensitivity response of the CEPAS lends for a particle size-resolved absorption measurement. Currently, much of the uncertainties in global climate models are related to black and brown carbon and how their optical properties are modified due to coating (lensing effect). Accumulation of coating material and particle mixing-state are dependent on coagulation and condensation, and both of these processes are driven in part by particle size.

iii) The relatively simple and straightforward design of the CEPAS allows it to be integrated into a compact and stand-alone sensor. Air quality sensors represent an emerging technology, which allows for a higher spatial resolution air quality monitoring. Black carbon is emitted in incomplete combustion, and urban areas entail multiple local point sources, such as vehicular exhaust emissions and biomass burning, in a relatively small area. Therefore, to make an accurate air quality assessment, a network of sensors, which are easy to deploy and operate (e.g. no intermittently replaced filter tapes), is required.

**Funding**

The work was supported by the Academy of Finland (Project numbers 326444 and 314364), Academy of Finland Flagship funding (grant no. 337552), and by the Academy of Finland Flagship Programme, Photonics Research and Innovation (PREIN), decision number: 320167.

**References**

Ajtai, T., Á Filep, M. Schnaiter, C. Linke, M. Vragel, Z. Bozóki, G. Szabó, and T. Leisner. 2010. "A Novel Multi−wavelength Photoacoustic Spectrometer for the Measurement of the UV–vis-NIR Spectral Absorption Coefficient of Atmospheric Aerosols." Journal of Aerosol Science 41 (11): 1020-1029. doi:10.1016/j.jaerosci.2010.07.008.

Arnott, W. P., H. Moosmüller, C. F. Rogers, T. Jin, and R. Bruch. 1999. "Photoacoustic Spectrometer for Measuring Light Absorption by Aerosol: Instrument Description." Atmospheric Environment 33 (17): 2845-2852. doi:10.1016/S1352-2310(98)00361-6.

Bond, T. C., S. J. Doherty, D. W. Fahey, P. M. Forster, T. Berntsen, B. J. DeAngelo, M. G. Flanner, et al. 2013. "Bounding the Role of Black Carbon in the Climate System: A Scientific Assessment." Journal of Geophysical Research: Atmospheres 118 (11): 5380-5552. doi:10.1002/jgrd.50171.

Bond, T. C., G. Habib, and R. W. Bergstrom. 2006. "Limitations in the Enhancement of Visible Light Absorption due to Mixing State." Journal of Geophysical Research: Atmospheres 111 (D20). doi:10.1029/2006JD007315.

Cao, Y., K. Liu, R. Wang, W. Chen, and X. Gao. 2021. "Three-Wavelength Measurement of Aerosol Absorption using a Multi-Resonator Coupled Photoacoustic Spectrometer." Optics Express 29 (2): 2258-2269. doi:10.1364/OE.412922.

Davies, N. W., M. I. Cotterell, C. Fox, K. Szpek, J. M. Haywood, and J. M. Langridge. 2018. "On the Accuracy of Aerosol Photoacoustic Spectrometer Calibrations using Absorption by Ozone." Atmospheric Measurement Techniques 11 (4): 2313-2324. doi:10.5194/amt-11-2313-2018.

Fischer, D. A. and Geoffrey D. Smith. 2018. "A Portable, Four-Wavelength, Single-Cell Photoacoustic Spectrometer for Ambient Aerosol Absorption." Aerosol Science and Technology 52 (4): 393-406. doi:10.1080/02786826.2017.1413231.

Järvi, L., H. Hannuniemi, T. Hussein, H. Junninen, P. P. Aalto, R. Hillamo, T. Mäkelä, P. Keronen, E. Siivola, and T. Vesala. 2009. "The Urban Measurement Station SMEAR III: Continuous Monitoring of Air Pollution and Surface–atmosphere Interactions in Helsinki, Finland." Boreal Environment Research 14 (Supplement A): 86-109.

Karhu, J., H. Philip, A. Baranov, R. Teissier, and T. Hieta. 2020. "Sub-Ppb Detection of Benzene using Cantilever-Enhanced Photoacoustic Spectroscopy with a Long-Wavelength Infrared Quantum Cascade Laser." Optics Letters 45 (21): 5962-5965. doi:10.1364/OL.405402.

Kauppinen, J., K. Wilcken, I. Kauppinen, and V. Koskinen. 2004. "High Sensitivity in Gas Analysis with Photoacoustic Detection." Microchemical Journal 76 (1): 151-159. doi:10.1016/j.microc.2003.11.007.

Kuusela, T. and J. Kauppinen. 2007. "Photoacoustic Gas Analysis using Interferometric Cantilever Microphone." Applied Spectroscopy Reviews 42 (5): 443-474. doi:10.1080/00102200701421755.

Lack, D. A., E. R. Lovejoy, T. Baynard, A. Pettersson, and A. R. Ravishankara. 2006. "Aerosol Absorption Measurement using Photoacoustic Spectroscopy: Sensitivity, Calibration, and Uncertainty Developments." Aerosol Science and Technology 40 (9): 697-708. doi:10.1080/02786820600803917.

Lack, D. A., M. S. Richardson, D. Law, J. M. Langridge, C. D. Cappa, R. J. McLaughlin, and D. M. Murphy. 2012. "Aircraft Instrument for Comprehensive Characterization of Aerosol Optical Properties, Part 2: Black and Brown Carbon Absorption and Absorption Enhancement Measured with Photo Acoustic Spectroscopy." Aerosol Science and Technology 46 (5): 555-568. doi:10.1080/02786826.2011.645955.

Luoma, K., J. V. Niemi, M. Aurela, P. L. Fung, A. Helin, T. Hussein, L. Kangas, A. Kousa, T. Rönkkö, and H. Timonen. 2021. "Spatiotemporal Variation and Trends in Equivalent Black Carbon in the Helsinki Metropolitan Area in Finland." Atmospheric Chemistry and Physics 21 (2): 1173-1189. doi:10.5194/acp-21-1173-2021.

Miklós, A., P. Hess, and Z. Bozóki. 2001. "Application of Acoustic Resonators in Photoacoustic Trace Gas Analysis and Metrology." Review of Scientific Instruments 72 (4): 1937-1955. doi:10.1063/1.1353198.

Nakayama, T., H. Suzuki, S. Kagamitani, Y. Ikeda, A. Uchiyama, and Y. Matsumi. 2015. "Characterization of a Three Wavelength Photoacoustic Soot Spectrometer (PASS-3) and a Photoacoustic Extinctiometer (PAX)." Journal of the Meteorological Society of Japan.Ser.II 93 (2): 285-308. doi:10.2151/jmsj.2015-016.

Peltola, J., T. Hieta, and M. Vainio. 2015. "Parts-Per-Trillion-Level Detection of Nitrogen Dioxide by Cantilever-Enhanced Photo-Acoustic Spectroscopy." Optics Letters 40 (13): 2933-2936. doi:10.1364/OL.40.002933.

Petzold, A. and R. Niessner. 1995. "Novel Design of a Resonant Photoacoustic Spectrophone for Elemental Carbon Mass Monitoring." Applied Physics Letters 66 (10): 1285-1287. doi:10.1063/1.113271.

Qian, Y., T. J. Yasunari, S. J. Doherty, M. G. Flanner, W. K. M. Lau, J. Ming, H. Wang, M. Wang, S. G. Warren, and R. Zhang. 2015. "Light-Absorbing Particles in Snow and Ice: Measurement and Modeling of Climatic and Hydrological Impact." Advances in Atmospheric Sciences 32 (1): 64-91. doi:10.1007/s00376-014-0010-0.

Rossi, J., J. Uotila, S. Sharma, T. Laurila, R. Teissier, A. Baranov, E. Ikonen, and M. Vainio. 2021. "Photoacoustic Characteristics of Carbon-Based Infrared Absorbers." Photoacoustics 23: 100265. doi: 10.1016/j.pacs.2021.100265.

Tomberg, T., M. Vainio, T. Hieta, and L. Halonen. 2018. "Sub-Parts-Per-Trillion Level Sensitivity in Trace Gas Detection by Cantilever-Enhanced Photo-Acoustic Spectroscopy." Scientific Reports 8 (1): 1-7. doi:10.1038/s41598-018-20087-9.

Virkkula, A., T. Mäkelä, R. Hillamo, T. Yli-Tuomi, A. Hirsikko, K. Hämeri, and I. K. Koponen. 2007. "A Simple Procedure for Correcting Loading Effects of Aethalometer Data." Journal of the Air & Waste Management Association 57 (10): 1214-1222. doi:10.3155/1047-3289.57.10.1214.

Wang, G., H. Yi, P. Hubert, A. Deguine, D. Petitprez, R. Maamary, E. Fertein, J. M. Rey, M. W. Sigrist, and W. Chen. 2017. "Filter-Free Measurements of Black Carbon Absorption using Photoacoustic Spectroscopy." Proc. SPIE 10111, Quantum Sensing and Nano Electronics and Photonics XIV 10111: 1011136. doi:10.1117/12.2251093.

Wiegand, J. R., L. D. Mathews, and G. D. Smith. 2014. "A UV–Vis Photoacoustic Spectrophotometer." Analytical Chemistry 86 (12): 6049-6056. doi:10.1021/ac501196u.

Xu, Y. and V. Ramanathan. 2017. "Well Below 2 C: Mitigation Strategies for Avoiding Dangerous to Catastrophic Climate Changes." Proceedings of the National Academy of Sciences 114 (39): 10315-10323. doi:10.1073/pnas.1618481114.

Yu, Z., G. Magoon, J. Assif, W. Brown, and R. Miake-Lye. 2019. "A Single-Pass RGB Differential Photoacoustic Spectrometer (RGB-DPAS) for Aerosol Absorption Measurement at 473, 532, and 671 nm." Aerosol Science and Technology 53 (1): 94-105. doi:10.1080/02786826.2018.1551611.

Supplementary material for:

# Cantilever-enhanced photoacoustic measurement of light-absorbing aerosols

Juho Karhu[a], Joel Kuula[b], Aki Virkkula[b], Hilkka Timonen[b], Markku Vainio[c,d], and Tuomas Hieta[e]

[a]Metrology Research Institute, Aalto University, Espoo, Finland; [b]Atmospheric Composition Research, Finnish Meteorological Institute, Helsinki, Finland; [c]Department of Chemistry, University of Helsinki, Helsinki, Finland; [d]Photonics Laboratory, Physics Unit, Tampere University, Tampere, Finland; [e]Gasera Ltd., Turku, Finland

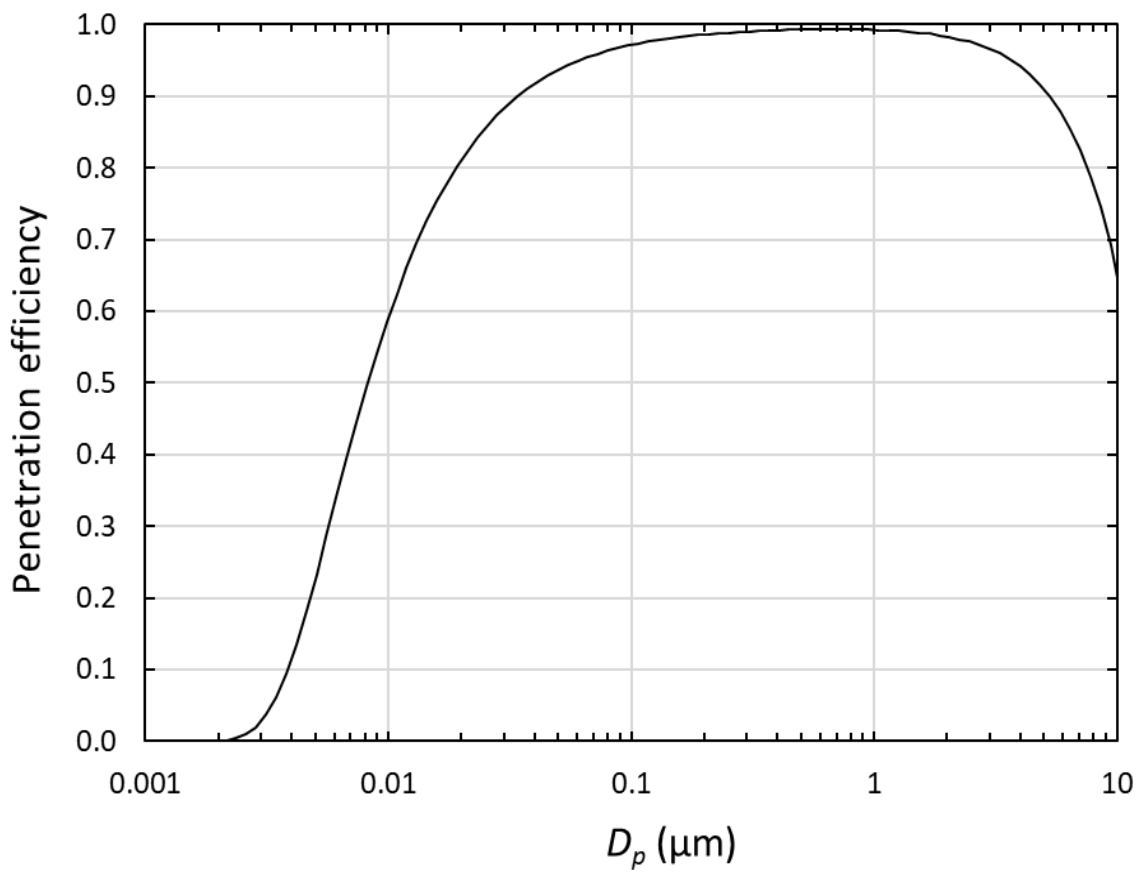

Figure S1: Particle diameter dependent penetration efficiency in the sampling line for outdoors air. The calculation takes into account the diffusion losses in 4 m of copper pipe and impaction losses in a 90° bend. The density of particles was set to 1.5 g cm$^{-3}$. The calculations were conducted according to formulas from (Hinds 1999) and (Baron and Willeke 2001).

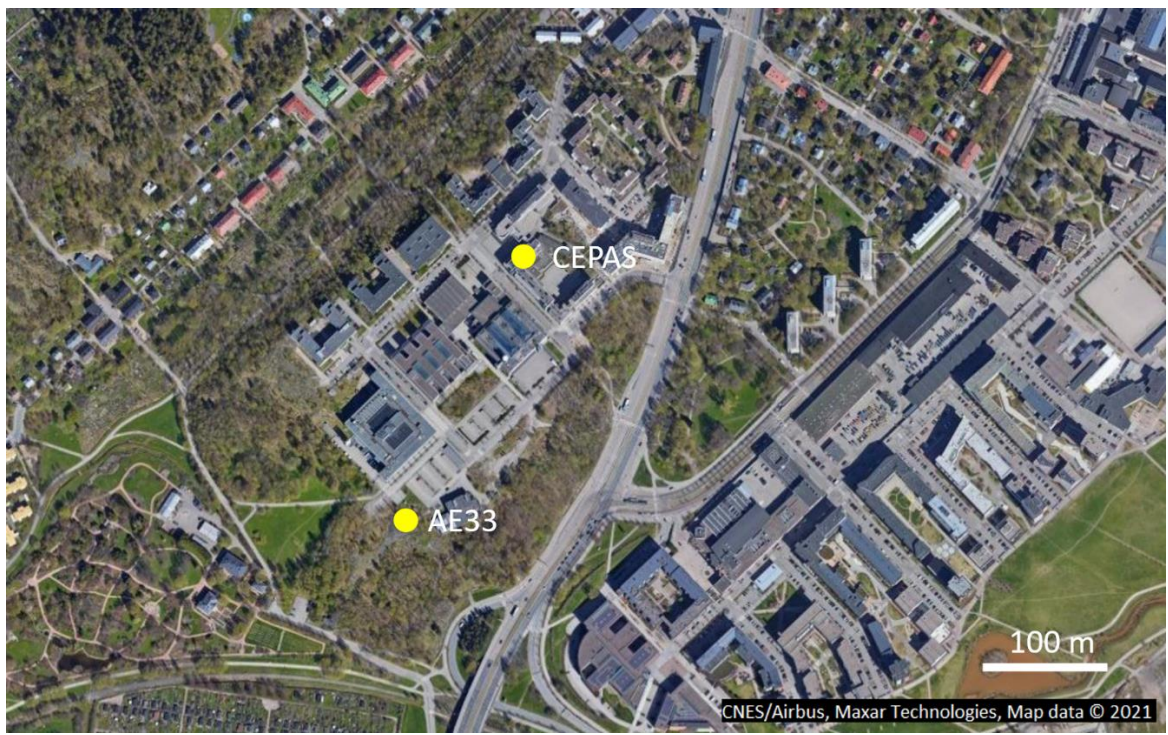

Figure S2. Map of the measurement locations. AE33 instrument was situated in the SMEARIII -station. It should be noted that the two measurement setups were located at different elevations. The AE33 instrument is located at ground level and the CEPAS measurement was sampled from 4$^{th}$ floor window.

**References**

Baron, P. A. and K. Willeke. 2001. Aerosol Measurement: Principles, Techniques, and Applications. 2nd ed. John Wiley & Sons.

Hinds, W. C. 1999. Aerosol Technology, Properties, Behaviour, and Measurement of Airborne Particles. 2nd ed. John Wiley & Sons.